\documentclass[prb,twocolumn,showpacs,preprintnumbers,superscriptaddress]{revtex4-1}
\usepackage[bookmarks, colorlinks=true, plainpages = true,linkcolor = blue, citecolor = blue, urlcolor = blue, filecolor =blue]{hyperref}

\usepackage{amsfonts}
\usepackage{graphicx}% Include figure files
\usepackage{amsmath}
\usepackage{amssymb}
\usepackage{latexsym}
\usepackage{psfrag}
\usepackage{dcolumn}% Align table columns on decimal point
\usepackage{datetime}
\usepackage{multirow}
\usepackage{subfigure}
\usepackage{amssymb,amsmath,rotate,color}
\usepackage{bm}% bold math
\begin{document}

\title{Copper atomic contacts exposed to water molecules}

%%%

%
\author{Firuz Demir}
\altaffiliation[Also at ]{the Department of Physics, Simon Fraser University, Burnaby, British Columbia, Canada V5A 1S6}
\affiliation{Department of Natural Sciences, University of South Carolina Beaufort, Bluffton, S.C.}
\author{Kevin Dean}
\affiliation{Physics Department,  Khalifa  University  of Science  and  Technology,  P.O.  Box  127788, Abu Dhabi, UAE}

\date{    {\color {red}         \today        }    }% It is always \today, today,  but any date may be explicitly specified

\begin{abstract}
Monatomic and molecular hydrogen and also oxygen, as well as water molecules and OH that are exposed to atomic copper in intimate contact have been studied theoretically using computational methods. The authors optimized moderately large structures of 
Cu/H/Cu,  
Cu/HCuH/Cu,
Cu/H$_2$/Cu,
Cu/H$_2$O/Cu, 
Cu/OH/Cu,
Cu/O/Cu,
Cu/O$_2$/Cu,
and calculated appropriate values for conductance and inelastic tunneling spectroscopy (IETS) properties of the contact junctions; elucidating them as being a possible outcome resulting from the exposure of copper electrodes to the atomic/molecular contaminant species.
Here we also demonstrate the IETS properties, by means of ab-initio calculations, that can determine the form of the junction geometries.
Furthermore, we identify the bonding geometries at the interfaces of the copper electrodes that directly  give rise to the specific IETS signatures that have been observed in recent experiments. Based on low-bias conductance and IETS calculations, for the specific case of water exposure of copper electrodes, it was concluded that a single hydrogen or a single oxygen atom bridging the copper electrodes is not responsible for the high conductance peak measurements. 
Regarding Model 4, where an individual water molecule is considered to be the bridging constituent, our computational results suggest that it has a relatively low probability of being an appropriate candidate.  Based upon current computational results, the two hydrogens in Model 3 appear to be in molecular form, although they still form a bond with the adjacent copper atoms.  Comparing computational with experimental results indicates that Model 3 is in acceptable agreement with available data.

\end{abstract}
\maketitle

% \section{Introduction}

Molecular nanowires, in which a single atom (or pair of atoms) or inorganic/organic molecules 
bond chemically with two metal electrodes simultaneously and forms a stable electrically conducting bridge between them, still continues to attract research attention because of the potential applications as single-molecule nanoelectronic devices. 
The excitation of molecular vibrations (phonons) by electrons passing through them  gives rise to quantized conductance steps. 
It is known that IETS experiments are able to detect these steps and the associated energies of the emitted phonons~\cite{Oxford_University_Press_Kirczenow}. 
However, due to experimental limitations, a single molecule located between the two electrodes is still not directly accessible to current scanning microprobes in order to measure atomic scale structure. 
A quantitative understanding of the electrical conduction through molecular wires, in order to gain additional control over their structures, still relies on a definitive determination of the bonding geometries at the molecule-electrode interfaces of the single molecule nanowires.

Based upon this incentive, we studied Cu/water/Cu junctions \cite{Yu_Li_PCCP2015}  and we are now revisiting these
systems in order to pursue further qualitative investigation of various fundamental quantum phenomena.
In this study we address issues related to hydrogen atoms and/or oxygen atoms,
that are considered to be impurities in copper wire systems, that may be incorporated unintentionally within the junctions during measurements.  
Whereas hydrogen bound to metals such as platinum and palladium has received considerable  attention, experimental evidence regarding hydrogen adsorbed into copper junctions has not been available to our knowledge. 
As a consequence, the detailed atomic arrangement is still not fully understood and conclusive evidence of, for example, whether hydrogen bonds directly to copper in atomic or molecular form is currently absent \cite{Frederiksen_JP_2007}.

\begin{figure}[t]
%\centerline{\includegraphics[width=1.0\linewidth]{ModelsConsidered.png}}
\centerline{\includegraphics[width=1.0\linewidth]{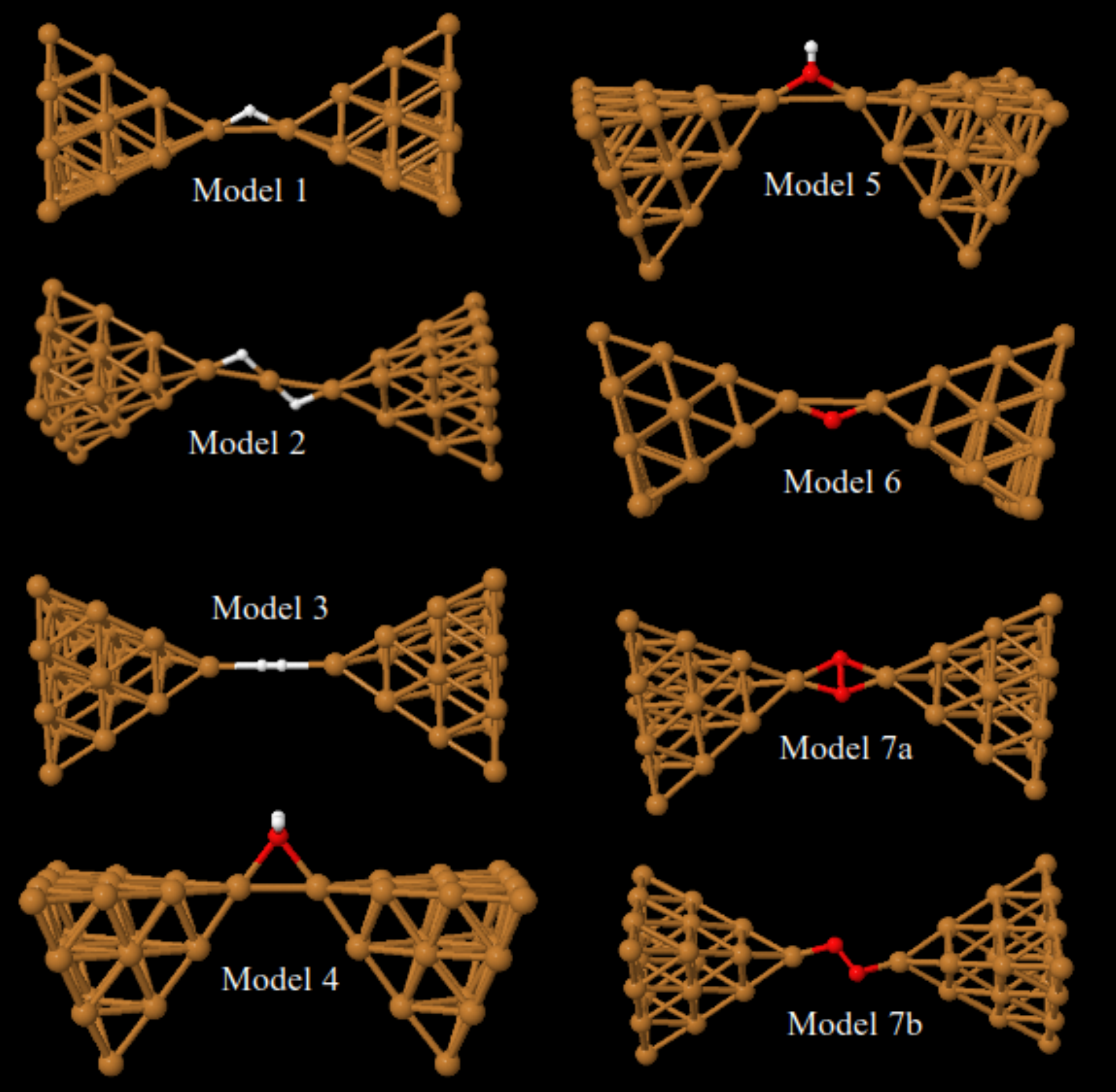}}
\caption{(Color online) Energetically favorable states of possible junction types.  
Copper, hydrogen, and oxygen atoms are brown, white, and red, respectively.
These models are the minimum energy configuration of the junctions which we began with to optimize.}
\label{ModelsConsidered}
\end{figure}

We benchmark our density functional theory (DFT) based calculations against available experimental data.   
Computational simulations reveal that the hydrogen and oxygen atoms of water molecules in the junction may also bond individually to the interface copper atoms within the junction at the nanowire leads.
This requires the water molecules (either bonding or non-bonding to the copper interface
atoms) must firstly lose their molecular bonds by means of unintentional tip-substrate contact. 
Our first principles calculations show this possibility as being highly plausible. Several examples of energetically favorable states of possible junction types, based on our calculations, are shown in Fig.~\ref{ModelsConsidered}. 
For purposes of atomic identification, the following color codes should be referred to: Brown – represents Cu atoms, White represents H atoms and Red represents Oxygen atoms.

%
%%%%%%%%%
%%%%%%%%%

%  \section{Calculations}  \label{Calculations}
Focused inelastic tunneling processes are sensitive to the copper/water interfaces, therefore it is necessary to calculate accurate equilibrium geometries of the complete molecule/copper electrode system. 
Unlike many previous theoretical studies that have been performed and reported by other groups, we do not freeze the geometry of the bulk copper crystal in order to perform relaxation only for the interfacial molecule, but also include the entire copper/water/copper system as an extended molecule in the optimization. We only constrain the distance of the outermost ends of our initial trial geometries. This limited restriction is very similar to the actual experimental setup and to our understanding it does not adversely affect the subsequent orientation of the remaining atoms in the extended molecule. Consequently, the individual components of the molecular structure can move freely in order to obtain the most computationally relaxed orientation.
Calculations were performed for a wide range of different bonding geometries by carrying out relaxations starting from different initial trial geometries. We carried out the relaxations at DFT level by using the PBE1PBE hybrid functional and Lanl2DZ basis sets with the GAUSSIAN'16 package~\cite{gaussian}. 
By using the same functional and basis sets, we proceeded with the optimized coordinates and  calculated the atomic displacements from equilibrium in the normal modes, as well as the corresponding frequencies and phonon energies. 
After identifying the normal modes of the extended molecule, having the maximum vibrational amplitudes specifically for the molecule at the interface and then calculating the frequencies of those modes (as outlined above), we determine which of these modes has the strongest IETS intensity. 
The normal modes with the strongest IETS intensities will give rise to the maximum conductance step height, upon applying the bias voltage across the extended molecule.

The theoretical approach that we adopt here is similar to that used in Refs.~\cite{Demir_Kirczenow_JCP_2011, Demir_Kirczenow_JCP_2012, Demir_Kirczenow_JCP_2012_2}. 
An in-depth discussion of the underlying theory and the relevant mathematical derivations that are required to calculate the IETS intensities perturbatively have been presented in 
Ref.~\cite{Demir_Kirczenow_JCP_2012}, so we will not repeat it here but  will focus attention on the details of our results.

The precise details regarding the methodology used  to couple the extended molecular structure to the macroscopic electron reservoirs and the calculation of the zero-bias tunneling conductances from the Landauer formula, are presented in 
Refs.~\cite{Demir_Kirczenow_JCP_2011, Demir_Kirczenow_JCP_2012, Demir_Kirczenow_JCP_2012_2}.  
These particular references also describe the solution of the Lippmann-Schwinger equation and evaluation of the appropriate Green's function by using semi-empirical extended H\"uckel theory \cite{yaehmop}, for the determination of the  elastic transmission amplitudes.

The zero-bias tunneling conductance values for molecular wires were calculated from the Landauer formula, 
\[  
g = g_0 \sum_{ij}    \vert
t_{ji}^{el}(\{{\bf 0}\}) \vert ^2 v_j/v_i~ 
\] with $g_0=2e^2/h$. 
Appropriate parametric values for the elastic transmission amplitudes were obtained, with detailed values being shown in Table~\ref{conductance}.  
The level of agreement between 
% these theoretical values and experimental values
the theoretical values and published experimental values 
is typical, with good correspondence to results from previously published studies, that were quoted as being: 
$0.2g_0$~\cite{Nakazumi_JPCC_118_2014}, 
$0.3g_0$~\cite{Yu_Li_JPCC_119_2015}
for Cu/H$_2$/Cu contacts, 
$0.1g_0$~\cite{Yu_Li_PCCP2015, Yu_Li_PCCP2017} 
for Cu/H$_2$O/Cu contacts, 
and
$0.1g_0$~\cite{Yu_Li_JPCC_120_2016} 
for Cu/O$_2$/Cu contacts. 
%
%
%
%%%%%%%%%%
%\begin{table}[b!]
%\caption{Energetically favorable states of possible junction types vs. their electric contact conductance.
%Conductance values shown in blue are either too high or too low, however, they are still within a typical experimental detection range and theoretical deviation. 
%}
%\label{conductance}
%%\centerline{\includegraphics[width=1.0\linewidth]{conductance.png}}
%\centerline{\includegraphics[width=1.0\linewidth]{conductance.pdf}}
%\end{table}
%%%%%%%%%%
%
%
%
\begin{table}[b!]
\caption{Energetically favorable states of possible junction types vs. their electric contact conductance.
Conductance values shown in blue are considered to be either too high or too low, however, they are still within a typical experimental detection range and theoretical deviation. 
}
\label{conductance}
\setlength{\tabcolsep}{0.3em} % for the horizontal padding
{\renewcommand{\arraystretch}{1.4}% for the vertical padding
{\fontfamily{cmss}\selectfont 
%{\small‎{
%{\scriptsize‎{
{\footnotesize‎{
\begin{tabular}{|c|l|c|c|}
\hline
\color{red}\textbf{Model} & 
\color{red}\textbf{Configuration}  & 
\color{red}\textbf{Junction Size ( \AA )} &  
\color{red}\textbf{Conductance (g$_0$)} 
\\ \hline
\textbf{1}     & \textbf{20Cu-H-20Cu}     & \textbf{2.92}                &   
\color{blue}\textbf{0.7844}           \\ \hline
\textbf{2}     & \textbf{20Cu-HCuH-20Cu}  & \textbf{5.11}             &  
\color{blue}\textbf{0.0124}           \\ \hline
\textbf{3}     & \textbf{20Cu-HH-20Cu}    & \textbf{4.99}                & 
\textbf{0.2716}           \\ \hline
\textbf{3}     & \textbf{20Cu-HH-20Cu}    & \textbf{4.97}                & 
\textbf{0.2834}           \\ \hline
\textbf{4}     & \textbf{20Cu-Water-20Cu} & \textbf{2.72}                & 
\textbf{0.3796}           \\ \hline
\textbf{4}     & \textbf{20Cu-Water-20Cu} & \textbf{2.77}                & 
\textbf{0.3142}           \\ \hline
\textbf{4}     & \textbf{20Cu-Water-20Cu} & \textbf{2.82}                & 
\textbf{0.2441}           \\ \hline
\textbf{4}     & \textbf{20Cu-Water-20Cu} & \textbf{2.86}                & 
\textbf{0.1966}           \\ \hline
\textbf{5}     & \textbf{20Cu-OH-20Cu}    & \textbf{3.29}                & 
\color{blue}\textbf{0.0019}           \\ \hline
\textbf{6}     & \textbf{20Cu-O-20Cu}     & \textbf{3.25}                & 
\color{blue}\textbf{0.0016}           \\ \hline
\textbf{7a}    & \textbf{20Cu-O2-20Cu}    & \textbf{3.75}                & 
\color{blue}\textbf{0.0057}           \\ \hline
\textbf{7b}    & \textbf{20Cu-O2-20Cu}    & \textbf{4.34}                &
\color{blue}\textbf{0.0114}           \\ \hline
\end{tabular}
}}
}
}
\end{table}

\begin{figure}[t!]
%\centerline{\includegraphics[width=1.0\linewidth]{00plot_IETS.png}}
\centerline{\includegraphics[width=1.0\linewidth]{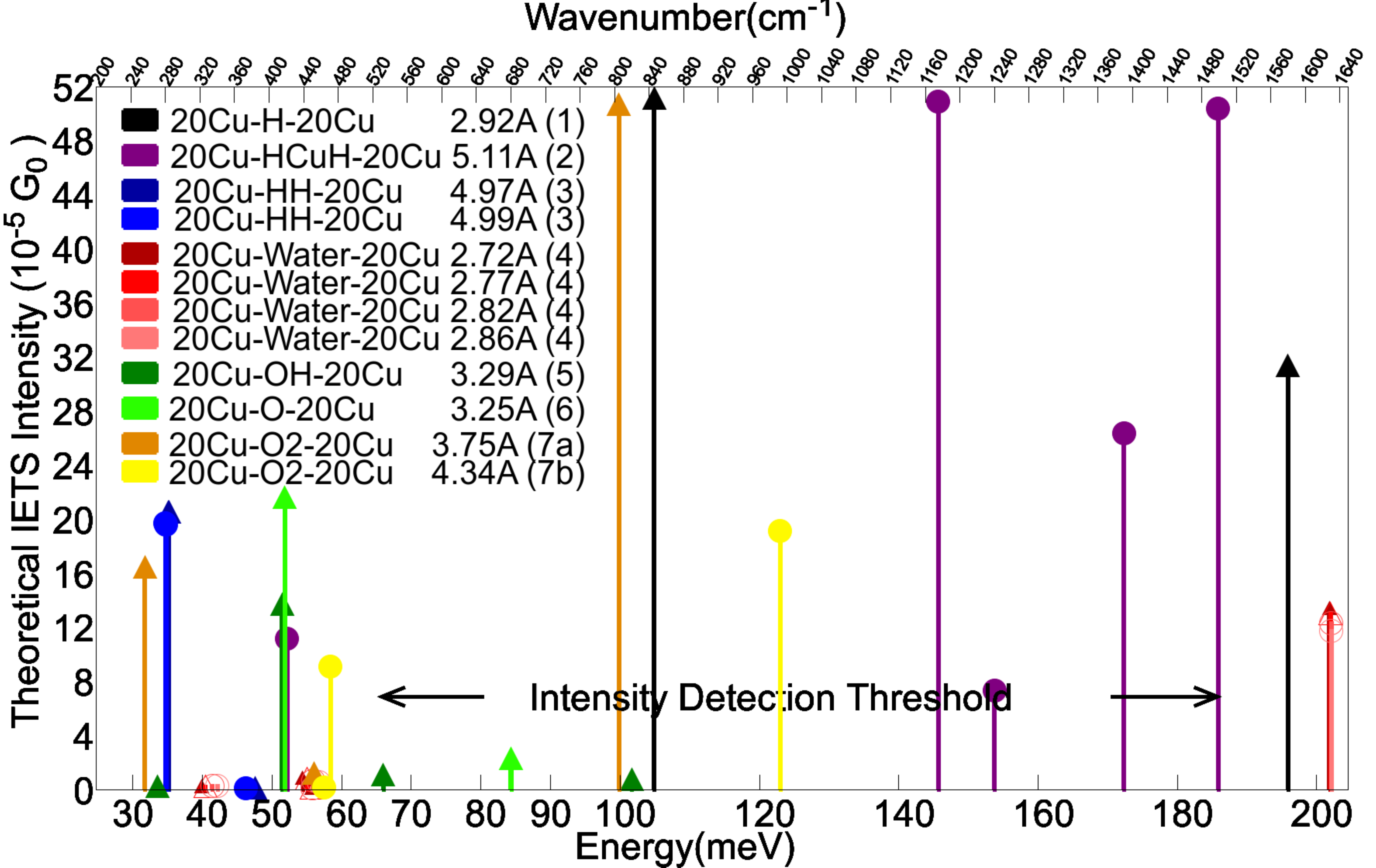}}
\caption{Comparison of electronic energies.}
\label{ElectronicEnergyCompare}
\end{figure}

We then calculated the vibrational modes, their respective frequencies and the IETS intensities for the extended molecules with various bonding geometries, 
as shown in Fig~\ref{ElectronicEnergyCompare}, 
and compared our results with the experimental data for the observed symmetrical peaks  in the dI/dV differential conductance curves for the following: Cu/H$_2$/Cu contacts~\cite{Nakazumi_JPCC_118_2014}, which is around $\pm 35$ meV, 
and Cu/O$_2$/Cu contacts~\cite{Yu_Li_JPCC_120_2016}, which is around $\pm 32$ meV and also around $\pm 60 $ meV
(average vibrational energy  69 meV).

These vibrational energies were determined at the DFT level. 
The scattering amplitudes ({\em i.e.} IETS intensities) for inelastic transmission of electrons through the molecular wire is sensitive to the change in the elastic 
amplitude for transmission through the wire when the atoms are displaced from their equilibrium positions. This occurs when the vibrational modes are excited with strong amplitudes.
Therefore, it is necessary for us to calculate accurate equilibrium geometries of the entire system, including both the molecule and the copper electrodes, and the phonon energies and atomic displacements from equilibrium for the vibrational normal modes of the whole system. 

Our plot (Fig~\ref{ElectronicEnergyCompare}) 
has prominent peaks near  $35$ meV for Cu/H$_2$/Cu models, 
and peaks near  $60$ meV and $120$ meV for Cu/O$_2$/Cu (Model 7b). 
It is notable that these peaks appear near  $30$ meV and $100$ meV for Cu/O$_2$/Cu for another model (specifically Model 7a) and it is apparent that upon further stretching the molecular geometry and the vibrational phonon modes consequently become slightly modified. 
%
%%%%%%%%%
%%%%%%%%%
%

As shown in Ref.~\cite{Demir_Kirczenow_JCP_2012}, our equation 
for IETS intensities, 
 %%%
\[ \label{omegaIntensityA} 
{\small 
\delta g_{\alpha}=  \frac{e^2}{2\pi \omega_{\alpha}} \lim_{A\to0}\sum_{ij} \frac{v_j}{v_i }   
\Big\vert \frac{ t_{ji}^{el}(\{A{\bf d}_{{n}\alpha}\})-t_{ji}^{el}(\{{\bf 0}\})} { A }
 \Big\vert ^2 
}
\]
%%%
states that the scattering amplitude for {\em inelastic} transmission of an electron through the molecular wire is directly proportional to the change in the {\em elastic} amplitude for transmission through the wire. 
When vibrational mode $\alpha$ is excited at low temperatures, atoms of the molecular wire are displaced from their equilibrium positions, which results in the calculated IETS intensities (i.e. the conductance step heights observed in experiments) $\delta g_{\alpha}$ being associated with the emission of phonons of mode $\alpha$.

All peaks, including the additional peak for Cu/O$_2$/Cu junctions near $120$ meV, are calculated at the zero-bias Fermi energy.  The values of the threshold bias at which the inelastic transmission occurs in the experiments are also low, therefore, as long as the atomic/molecular junctions remain stable the peaks with strong amplitudes should be readily detected in experiments.

% \section{Results}  \label{Results}

We optimized the model structures shown in Fig.~\ref{ModelsConsidered} and performed accurate calculations of the atomic displacements from equilibrium for these atoms in the respective vibrational modes.  
By comparing the calculated conductance and inelastic intensity of each model, we were able to consider several possibilities.

One mechanism that was considered is for hydrogen and oxygen atoms in water molecules to bond directly to copper atoms in the junction. 
This possibility may not be expected in terms of chemistry alone, however, the types of experiments due to their mechanical settings may
provide the possibility that water molecules could lose their usual molecular bonds and subsequently form individual bonds with the nearest-neighbor copper atoms in the junction.

When hydrogen and/or oxygen atoms form bonds in the junction, they act analogously as impurities within the metalic crystal structure and consequently lower the conductance. This conductance change is not paramount for some bonding types; therefore, as a result this change may not be easily detectable in experiments and the signatures of their bonding geometries will not necessarily be readily identifiable.

With a single H atom bonding simultaneously to a pair of interface copper atoms (see Fig.~\ref{ModelsConsidered} Model 1), we have calculated strong inelastic intensities in the vicinity of 105 and 196 meV.
%  845.59	104.85	
% 1580.30	195.94
% 
In the case of 2 hydrogen atoms (where the hydrogens are not specifically thought to be in molecular form; see Fig.~\ref{ModelsConsidered} Model 2) we calculate strong IETS intensity signals in the vicinity of 52, 146, 154, 173, and 187 meV 
(146 and 187 meV are exceptionally strong).
%   422.8814	52.43	
%   1176.77	145.91	
%   1241.262	153.91	
%   1397.961	173.34	
%   1506.037	186.74

When the hydrogens are thought to be in molecular form but still making two bonds with the internal copper atoms (see Fig.~\ref{ModelsConsidered} Model 3), 
these modes are close to 35 and 48 meV.
%  283.47	35.15	
%  284.05	35.22	
%  383.89	47.6		
%  3895.39	483.00
For the mode near 48 meV we calculate a very weak IETS intensity, and therefore it is considered to be below the threshold for experimental detection and so consequently, we don't anticipate it to be successfully measured in experiments. 

For Model 5 (see Fig.~\ref{ModelsConsidered}) where oxygen is simultaneously bridging two interface copper atoms while bonding to a single hydrogen, we have modes near  
34, 51, 66 and 102 meV.
%  270.68	33.56	
%  414.66	51.42	
%  532.17	65.99	
%  819.73	101.64	
%  3813.24	472.81
The mode near 51 meV is the only one that is regarded as being above our detection threshold region. 
However, Model 2, Model 6 and Model 7b also have acceptable modes in the vicinity of 55 and 60 meV, and display an experimentally detectable strong IETS signal at this level. However, this should not automatically be assigned to Model 5.
The elastic scattering calculations for these modes are not sufficiently different so as to specifically identify the appropriate modes and consequently identify their junction types, by only comparing their calculated conductance values. 
With the help of IETS, the interface geometries can still be confidently identified with the following procedure: If a junction has a mode measured between 50 meV and 60 meV but no IETS signal is detected near 120 meV and/or above 145 meV for the same junction, it is less likely that the junction type is as calculated for Model 7b and/or Model 2.

There are two high IETS intensities, 
near 105 meV for Model 1
%  845.59	104.85	
% 1580.30	195.94
and near 100 meV for Model 7a.
%  255.99	31.74		
%  451.55	55.99	
%  804.53	99.76
The above procedure may also be applied here in order to assign the signal to a specific junction type when measured in experiments. 
Model 7a has another mode near 32 meV with high IETS intensity while Model 1 has no mode near this energy. 
Similarly, Model 1 has another mode near 195 meV with high IETS intensity while Model 7a has no mode near this particular energy.

When water molecules are studied as they are exposed to copper electrodes in intimate contact, it is considered to be highly probably that they will form a bridge and we can confidently assign the rapid conductance drop to this fact. However, the calculated conductance frequently appears to be too high for Cu/H$_2$O/Cu junctions. If the molecular structure is not under high stress (unlike our Model 4), the conductance fluctuates between ~0.5~g$_0$ and ~0.8~g$_0$. 
Model 4 is held fixed from the outer ends,  during our systematic treatment in this study, and the remaining extended molecule is therefore optimized within this constrained degree of freedom. 
For this model our conductance calculation is near 0.2~g$_0$, which is in excellent  correspondence to results from previously published studies, 
$0.1g_0$~\cite{Yu_Li_PCCP2015, Yu_Li_PCCP2017} 
for Cu/H$_2$O/Cu contacts. 
However, it is observed that Model 4 has no vibrational mode within our detection threshold criterion, until 202 meV. The vibrational modes near 42 and 56 meV have noticeably weak IETS and fall well below 
% the 
our estimated
threshold for experimental detection. 
%  340.26	42.19	
%  449.69	55.76	
%  457.27	56.70	
%  1630.47	202.17	
%  3667.22	454.71	
%  3832.92	475.25
Therefore, we think that the water molecules bridging the copper electrodes are not intact and are consequently not directly responsible for the measured conductance values in experiments. We suggest further experimental investigation should clarify this situation.

%\section{Conclusions}
 
In conclusion: We have studied  water molecules that are exposed to atomic copper in intimate contact. 
We computationally optimized moderately large structures of 
Cu/H/Cu,  
Cu/HCuH/Cu,
Cu/H$_2$/Cu,
Cu/H$_2$O/Cu, 
Cu/OH/Cu,
Cu/O/Cu,
Cu/O$_2$/Cu,
and calculated appropriate values for the conductance and IETS properties of the contact junctions and considered them as being a highly probable outcome resulting from the water exposure of copper electrodes. 
We have successfully demonstrated the IETS properties, by means of ab-initio calculations and identified the hypothesized bonding geometries at the interfaces of the copper electrodes. 

Based on low-bias conductance calculations alone, for the case of 
monatomic and molecular hydrogen and also oxygen, as well as water molecule
exposure of copper electrodes, we found the  (apparently) relatively low possibility of hydrogen or oxygen  bridging of the copper electrodes, as in 
Models 1, 2, 5, 6, 7a, 7b and therefore giving rise to the high conductance experimental peaks. When a water molecule is modeled as the bridging constituent, the conductance calculations remain in good agreement with the recently available experimental data~\cite{Yu_Li_PCCP2015, Yu_Li_PCCP2017}. 
However, it should be noted that based solely on our IETS calculations this possibility actually becomes weakened due to undetectably weak IETS near 43 meV and 57 meV, but detectably strong IETS near 202 meV.

Our results have demonstrated that the conductance values of molecular junctions with Cu electrodes are apparently insensitive to the presence of hydrogens and thus it is difficult to unequivocally determine directly if an atomic copper wire contains hydrogen impurities based on conductance measurements alone, without resorting to measuring the inelastic signals. 
 In copper junctions with different impurities there are also similar inelastic signals observed below  approximately 50 to 60 meV, which are in close proximity. 
Therefore, it becomes a big challenge to specifically identify the kind of 
specific
impurity unless we investigate the inelastic signals at much higher phonon energies. In the case of a single H atom (Fig.~\ref{ModelsConsidered} Model 1) our calculations predict signals around 105 and 196 meV.  
Our results also show that there is no mode below 200 meV with strong IETS in the case of Model 4 (Fig.~\ref{ModelsConsidered}).
When hydrogens are in molecular form but still making two bonds with the internal copper atoms (as in Fig.~\ref{ModelsConsidered} Model 3), we calculate high IETS near 35 meV. It is of interest to note that Model 7a also has a mode with high IETS close to 35 meV.  
However, the weak conductance calculation of Model 7a makes it a less likely candidate for forming this type of junction in Cu/water/Cu experiments while Model 3 still remains within good agreement with the experiments, based on both conductance and IETS calculations.
We find that the water molecules bridging copper electrodes are not necessarily intact in Cu/water/Cu experiments and they 
are  
responsible for the measured conductance values in experiments. 
{\em i.e.} water molecules can associate via weak hydrogen bonds and it is considered quite probable that these bonds will dissociate to break them during the experimental procedures.

The analyses we adopted here will lead to the identification of the specific type of impurity as well as its configuration in the junction.
Consequently, we would like to encourage further experimental investigation.

%  \section{Acknowledgments}
%
The authors wish to acknowledge the invaluable contribution of the high-performance computing facilities at WestGrid and Compute Canada, and also the HPC cluster at the Masdar Institute in the UAE,  to the results of this research.
We also thank G. Kirczenow, A. Saffarzadeh and M. Kiguchi for helpful discussions.


\begin{thebibliography}{99}

\bibitem {Oxford_University_Press_Kirczenow}
For a review see G. Kirczenow, ``The Oxford Handbook of Nanoscience and Technology: Volume 1: Basic Aspects,"
(Oxford University Press, Inc., U.K., 2010) Chap. 4.

\bibitem{Yu_Li_PCCP2015}
Y. Li, F. Demir, S. Kaneko, S. Fujii, T. Nishino, A. Saffarzadeh, G. Kirczenow, and M. Kiguchi, 
Phys. Chem. Chem. Phys. {\bf 17}, 32436-32442 (2015). 


\bibitem{Frederiksen_JP_2007}
T. Frederiksen, M. Paulsson, and M. Brandbyge, 
J. Phys.: Conference Series  {\bf 61}, 312-316 (2007).  
 
\bibitem{Demir_Kirczenow_JCP_2011}
F. Demir and G. Kirczenow, J. Chem. Phys. {\bf 134}, 121103 (2011).
 
\bibitem{Demir_Kirczenow_JCP_2012}
F. Demir and G. Kirczenow, J. Chem. Phys. {\bf 136}, 014703 (2012).
 
\bibitem{Demir_Kirczenow_JCP_2012_2}
F. Demir and G. Kirczenow, J. Chem. Phys. {\bf 137}, 094703 (2012).


\bibitem{gaussian}
M. J. Frisch, G. W. Trucks, H. B. Schlegel, G. E. Scuseria, M. A.
Robb, J. R. Cheeseman, G. Scalmani, V. Barone, B. Mennucci, G. A.
Petersson \textit{et al.}, Gaussian 16 Revision: A.03, 
Gaussian, Inc., Wallingford, CT, 2016.


\bibitem{yaehmop}The version of extended H\"{u}ckel theory used 
was that of J. H. Ammeter, H.-B. B\"{u}rgi, J. C. Thibeault, and R. 
Hoffman, J. Am. Chem. Soc. {\bf 100}, 3686 (1978), as implemented in
the YAEHMOP numerical package by G. A. Landrum and W. V. Glassey
(Source-Forge, Fremont, California, 2001).


\bibitem{Nakazumi_JPCC_118_2014} T. Nakazumi, S. Kaneko, and M. Kiguchi,     J. Phys. Chem. C {\bf 118}, 7489-7493 (2014).

\bibitem{Yu_Li_JPCC_119_2015} Y. Li, S. Kaneko, S. Fujii, and M. Kiguchi,     J. Phys. Chem. C {\bf 119}, 19143-19148 (2015).

\bibitem{Yu_Li_JPCC_120_2016} Y. Li, S. Kaneko, Y. Komoto, S. Fujii, T. Nishino, and M. Kiguchi, J. Phys. Chem. C {\bf 120}, 16254-16258 (2016).

 

\bibitem{Yu_Li_PCCP2017} Y. Li, S. Kaneko, S. Fujii, T. Nishino, and M. Kiguch, 
Phys. Chem. Chem. Phys. {\bf 19}, 4673 (2017). 








\end{thebibliography}
\end{document}